\documentclass[12pt]{JHEP3}
\usepackage{epsfig} 
\usepackage{amsmath} 
\usepackage{amssymb}
\usepackage{graphicx}
\usepackage{float}
\restylefloat{table}

\large
\newcommand{\beq}{\begin{equation}}
\newcommand{\eeq}{\end{equation}}
\newcommand{\ba}{\begin{array}}
\newcommand{\ea}{\end{array}}
\newcommand{\bea}{\begin{eqnarray}}
\newcommand{\eea}{\end{eqnarray}}
\def\nn{\nonumber}
\def\bd{B_d^0}
\def\bs{B_s^0}
\def\bdbar{{\bar B}_d^0}
\def\bsbar{{\bar B}_s^0}
\def\dub{\delta_{ub}}
\def\dtpd{\delta_{t'd}}
\def\dtps{\delta_{t's}}

\def\ctl       {\ensuremath{\cos{\theta_l}}}
\def\stl       {\ensuremath{\sin{\theta_l}}}
\def\ctk       {\ensuremath{\cos{\theta_K}}}
\def\sstk       {\ensuremath{\sin^2{\theta_K}}}
\def\cstk       {\ensuremath{\cos^2{\theta_K}}}
\def\sstl       {\ensuremath{\cos2{\theta_l}}}
\def\sstl       {\ensuremath{\sin^2{\theta_l}}}

\def\cttl       {\ensuremath{\cos2{\theta_l}}}
\def\sttl       {\ensuremath{\sin2{\theta_l}}}

\def\sttk       {\ensuremath{\sin2{\theta_K}}}
\def\cp       {\ensuremath{\cos{\phi}}}
\def\sp       {\ensuremath{\sin{\phi}}}
\def\stp       {\ensuremath{\sin2{\phi}}}
\def\ctp       {\ensuremath{\cos2{\phi}}}
\newcommand{\lqcd}{\ensuremath{\Lambda_{\mathrm{QCD}}}}
\def\gsim{{~\raise.15em\hbox{$>$}\kern-.85em
          \lower.35em\hbox{$\sim$}~}}
\def\lsim{{~\raise.15em\hbox{$<$}\kern-.85em
          \lower.35em\hbox{$\sim$}~}}

\newcommand{\gev}{\ensuremath{\mathrm{\,Ge\kern -0.1em V}}}

\title{New-physics signals of a model with a vector-singlet up-type quark}

\author{Ashutosh Kumar Alok and Subhashish Banerjee \\
Indian Institute of Technology Jodhpur, Jodhpur 342011, India \\
E-mail: \email{akalok@iitj.ac.in}, \email{subhashish@iitj.ac.in}}

\author{Dinesh Kumar$^{a,b}$ and S. Uma Sankar$^a$ \\
{$^a$}Indian Institute of Technology Bombay, Mumbai 400076, India \\
{$^b$}Department of Physics,
University of Rajasthan, Jaipur 302004, India \\
E-mail: \email{dinesh@phy.iitb.ac.in}, \email{uma@phy.iitb.ac.in}}

\author{David London \\
Physique des Particules, Universit\'e de Montr\'eal, \\ 
C.P. 6128, succ. centre-ville, Montr\'eal, QC, Canada H3C 3J7 \\
E-mail: \email{london@lps.umontreal.ca}}

\abstract{The VuQ model involves the addition of a vector isosinglet
  up-type quark to the standard model. In this model the full CKM
  quark mixing matrix is $4 \times 3$. Using present flavor-physics
  data, we perform a fit to this full CKM matrix, looking for signals
  of new physics (NP).  We find that the VuQ model is very strongly
  constrained.  There are no hints of NP in the CKM matrix, and any
  VuQ contributions to loop-level flavor-changing $b \to s$, $b \to d$
  and $s \to d$ transitions are very small. There can be significant
  enhancements of the branching ratios of the flavor-changing decays
  $t \to u Z$ and $t \to c Z$, but these are still below present
  detection levels.}


\preprint{UdeM-GPP-TH-15-241}

\begin{document} 

%%%%%%%%%%%%%%%%%%%%%%%%%%%%%%%%%%%%%%%
\section{\bf Introduction}
%%%%%%%%%%%%%%%%%%%%%%%%%%%%%%%%%%%%%%%

The standard model (SM) includes three generations of fermions. In
particular, there are three down-type quarks ($Q_{em} = -1/3$: $d$,
$s$, $b$) and three up-type quarks ($Q_{em} = 2/3$: $u$, $c$, $t$).
All quarks with a given charge mix, so that there is a $W$ coupling
between each down-type and up-type quark. These couplings are
tabulated in the Cabibbo-Kobayashi-Maskawa (CKM) matrix.

Now, there is no {\it a-priori} reason for there to be only three
down-type and three up-type quarks. Indeed, many models of physics
beyond the SM include new, exotic quarks. The simplest of these
consider a fourth generation of quarks (denoted SM4), a vector
isosinglet down-type quark $b'$ (denoted VdQ; both $b'_L$ and $b'_R$
have weak isospin $I=0$), or a vector isosinglet up-type quark $t'$
(denoted VuQ; both $t'_L$ and $t'_R$ have weak isospin $I=0$).

There are two distinct ways to look for signals of such new physics
(NP). The first is via direct searches at colliders. To date, no
signals of exotic quarks have been observed; the limits on the masses
of these quarks depend on the assumptions about how they decay. Some
recent results are (this is not exhaustive) $m_{b',t'} > 685$ GeV
(95\% C.L.)  for the SM4 model \cite{CMSSM4}, $m_{b'} \gsim 450$ GeV
for the VdQ model \cite{CMSVdQ}, and $m_{t'} >$ 687-782 GeV (95\%
C.L.)  for the VuQ model \cite{CMSVuQ}.

Second, one can look for indirect signals of the exotic quarks through
their loop-level contributions to various processes. In fact, it is
possible to simultaneously consider all such loop-level effects. This
is done as follows. Most of these NP effects are charged-current
interactions, which involve the CKM matrix.  In the SM, the CKM matrix
is $3 \times 3$ and unitary. As such, it is parametrized by four
parameters. However, in all NP models the full mixing matrix is larger
than $3 \times 3$, so its parametrization requires additional
parameters.  The idea is then to perform a fit to the full CKM matrix
using all the data. A signal of the NP will be the non-unitarity of
the $3 \times 3$ CKM matrix. That is, some of the NP parameters will
be found to be nonzero.

At first glance, the analysis to search for NP is the same for all
three models. First, in all cases the parametrization of the full CKM
matrix has four SM and five NP parameters.  Second, one uses the same
flavor-physics data to perform a combined fit to these parameters.
This yields the best-fit values of all the parameters, and indicates
whether any of the NP parameters can be nonzero. However, the key
point is that the contributions to the flavor-physics observables are
model-dependent. That is, the effects on the observables vary from
model to model, so that the analyses are {\it not} the same for the
three models. The SM4 and VdQ models were examined in Refs.~\cite{SM4}
and \cite{dVQM}, respectively. In the present paper we consider the
VuQ model \cite{AgSaav,BBN}, in which the full CKM matrix is $4 \times
3$.

For the fit, in addition to the six directly-measured magnitudes of
CKM matrix elements, we include flavor-physics observables that have
small hadronic uncertainties:
(i) $\epsilon_K$ from CP violation in $K_L \to \pi \pi$, 
(ii) the branching fractions of $K^+ \to \pi^+ \nu \bar{\nu} $ and $K_L \to \mu^+ \mu^-$,
(iii) $R_{b}$ and $A_b$ from $Z\to b \bar{b}$, 
(iv) $\bs$-$\bsbar$ and $\bd$-$\bdbar$ mixing, 
(v) the time-dependent indirect $CP$ asymmetries in $\bd \to J/\psi\, K_S$ and $\bs \to J/\psi\, \phi$, 
(vi) the measurement of the CP-violating angle $\gamma$ of the
unitarity triangle from tree-level decays,
(vii) the branching ratios of the inclusive decays $B \to X_s l^+
l^-$ and $B \to X_s\, \gamma$, and of the exclusive decay $B \to K
\mu^+ \mu^-$,
(viii) many observables in $B \to K^* \mu^+ \mu^-$,
(ix) the branching ratios of $\bs \to \mu^+ \mu^-$, $\bd \to \mu^+ \mu^-$ and $B^+ \to \tau^+ \nu_{\tau}$, 
(x) the like-sign dimuon charge asymmetry $A^b_{SL}$, 
(xi) the oblique parameters $S$ and $T$. 
The fit is carried out for $m_{t'}=800$ GeV and $1200$ GeV.

In the VuQ model, the $t'_L$ can mix with the $u_L$, $c_L$ and
$t_L$. However, because the $t'_L$ and $\{u_L, c_L, t_L\}$ have
different values of $I_{3L}$ ($I_{3L} = 0$ for $t'_L$, $I_{3L} =
\frac12$ for $\{u_L, c_L, t_L\}$), this mixing will induce tree-level
$Z$-mediated flavor-changing neutral currents (FCNC's) among the SM
quarks. In particular, this means that $D^0$-${\bar D}^0$ mixing
occurs at tree level. Thus, in principle there can be constraints from
the experimental measurement of this mixing. Now, in the SM, this
mixing is due to a box diagram with internal $d$, $s$ and $b$
quarks. The $b$ contribution suffers a significant CKM suppression of
$O(\lambda^8)$, so that $D^0$-${\bar D}^0$ mixing is dominated by the
contributions of the internal $d$ and $s$ quarks. Because these quarks
are light, there can be large long-distance (LD) contributions to the
mixing. At present, there is no definitive estimate of these LD
effects.  As a result, we do not have an accurate prediction of the
value of $D^0$-${\bar D}^0$ mixing within the SM, so that this
measurement cannot be incorporated into the fit.

Once the fit has been performed, we can then make predictions for
other quantities that are expected to be affected by the $t'$ quark,
while still being consistent with the above measurements. We examine
the following observables:
(i) the branching fraction of $K_L \to \pi^0 \nu \bar{\nu}$, 
(ii) the branching fraction of $B \to X_s \nu \bar{\nu}$, 
(iii) $D^0$-${\bar D}^0$ mixing and the branching fraction of $D^0 \to \mu^+ \mu^-$, and
(iv) the branching fraction of $t\to q Z$ ($q=u,c$).

The paper is organized as follows. In Sec.~\ref{ckm4}, we define the
CKM parametrization and discuss the measurements used in the $\chi^2$
fit. The results of the fit are presented in Sec.~\ref{res-fit}. Given
these results, we calculate the possible effects of the VuQ model on
several other flavor observables in Sec.~\ref{pred:flavor}.
Sec.~\ref{concl} summarizes the results.

%%%%%%%%%%%%%%%%%%%%%%%%%%%%%%%%%%%%%%%
\section{\bf Constraints on the CKM Matrix}
\label{ckm4}
%%%%%%%%%%%%%%%%%%%%%%%%%%%%%%%%%%%%%%%

In the VdQ model the CKM matrix is $3 \times 4$. It was shown in
Ref.~\cite{DK} that this is the upper $3 \times 4$ submatrix of the $4
\times 4$ SM4 CKM matrix, denoted CKM4. Now, there are many
parametrizations of CKM4. For the VdQ model, it is best to choose one
in which the new matrix elements $V_{ub'}$, $V_{cb'}$ and $V_{tb'}$
take simple forms. With this in mind, the Dighe-Kim parametrization of
Refs.~\cite{DK,Alok:2008dj} was used in Ref.~\cite{dVQM}.

The logic is similar for the VuQ model. In this model the CKM matrix
is $4 \times 3$:
\beq
V_{\rm VuQ}=\left(\begin{array}{ccc}
V_{ud}& V_{us}&V_{ub}\\ V_{cd}&V_{cs}& V_{cb}\\
V_{td}&V_{ts}&V_{tb}\\
V_{t'd}&V_{t's}&V_{t'b}\\
\end{array}\right)~.
\eeq
$V_{\rm VuQ}$ is the left-hand $4 \times 3$ submatrix of CKM4.  Here
it is best to choose a parametrization of CKM4 in which the new matrix
elements $V_{t'd}$, $V_{t's}$ and $V_{t'b}$ take simple forms. We use
the Hou-Soni-Steger parametrization \cite{HSS,NandiSoni}.  Here,
\beq
\begin{tabular}{lll}
$V_{us}  \equiv  \lambda$ , &
$V_{cb}  \equiv A \lambda^2$ , &
$V_{ub}  \equiv  A \lambda^3 C e^{-i\dub}$ , \\
$V_{t'd} \equiv  -P \lambda^3 e^{i\dtpd}$ , &
$V_{t's}  \equiv -Q \lambda^2 e^{i\dtps}$ , &
$V_{t'b}  \equiv -r \lambda $ ~,
\end{tabular}
\label{ckm4tab}
\eeq
where $\lambda$ is the sine of the Cabibbo angle. There are four SM
parameters ($\lambda$, $A$, $C$, $\dub$) and five NP parameters ($P$,
$Q$, $r$, $\dtpd$, $\dtps$). Of the remaining six CKM matrix elements,
$V_{ud}$, $V_{cd}$ and $V_{cs}$ retain their SM parametrizations:
\beq
V_{ud} = 1 - \frac{\lambda^2}{2} ~,~~
V_{cd}  =  -\lambda  ~,~~
V_{cs}  =  1 - \frac{\lambda^2}{2} ~,
\label{vud} 
\eeq
but $V_{td}$, $V_{ts}$ and $V_{tb}$ are modified:
\bea
V_{td} & = & A \lambda^3 \left( 1 - C e^{i\dub} \right) - P r \lambda^4 e^{i\dtpd} + \frac12 A C \lambda^5 e^{i\dub} ~, \nn\\
V_{ts} & = & -A \lambda^2 - Q r \lambda^3 e^{i\dtps} + A \lambda^4 \left( \frac12 - C e^{i\dub} \right) ~, \nn\\
V_{tb} & = & 1 - \frac12 r^2 \lambda^2 ~.
\label{vtb}
\eea

In the limit $P=Q=r=0$, only the elements present in the $3\times 3$
CKM matrix retain nontrivial values, and the above expansion
corresponds to the Wolfenstein parametrization \cite{Wolfenstein} with
$C= \sqrt{\rho^2 + \eta^2}$ and $\dub = \tan^{-1}(\eta/\rho)$. In this
limit, $V_{tb} = 1$. In the VuQ model, $r$ can be nonzero, leading to
a deviation of $V_{tb}$ from 1.

For the fit, we consider all observables that can constrain the
parameters of the CKM matrix. The total $\chi^2$ is written as a
function of these parameters, and their best-fit values are those that
minimize this $\chi^2$ function. The total $\chi^2$ function is
defined as
\bea
\chi^2_{\rm total} & = & 
\chi^2_{\rm CKM} 
+ \chi^2_{|\epsilon_K|} 
+ \chi^2_{K \to \pi^+ \nu \bar{\nu}} 
+ \chi^2_{K_L \to \mu^+ \mu^-}
+ \chi^2_{Z\to b{\bar b}} 
+ \chi^2_{\bd} 
+ \chi^2_{M_R}
+ \chi^2_{\sin 2\beta} \nn\\
&& \hskip2truemm 
+~\chi^2_{\sin 2\beta_s} 
+ \chi^2_{\gamma} 
+ \chi^2_{B \to X_s\, l^+ \,l^-} 
+ \chi^2_{B \to X_s\, \gamma} 
+ \chi^2_{B \to K\, \mu^+ \,\mu^-} 
+ \chi^2_{B  \to K^* \, \mu^+ \,\mu^-} \nn\\ 
&& \hskip2truemm 
+~\chi^2_{B^+ \to \pi^+\, \mu^+\, \mu^-}
+\chi^2_{B_q \to \mu^+ \mu^-}
+ \chi^2_{B \to \tau \,\nu}
+ \chi^2_{A^b_{SL}}
+ \chi^2_{\rm Oblique}  ~.
\eea
In our analysis, the $\chi^2$ of an observable $A$ whose measured
value is $(A_{exp}^c \pm A_{exp}^{err})$ is defined as
\beq
\chi^2_A = \left( \frac{A - A_{exp}^c}{A_{exp}^{err}} \right)^2 ~.
\eeq
In the following subsections, we discuss the various experimental
measurements used in the fit, and give their individual contributions
to $\chi^2_{\rm total}$.
 
The current experimental values for the 68 flavor-physics observables
enumerated in the introduction are listed in Tables \ref{tab1} and
\ref{bkstar}. The theoretical expressions for these observables
require additional inputs in the form of decay constants, bag
parameters, QCD corrections and other parameters. These are listed in
Table \ref{tab3}.

%%%%%%%%%%%%%%%%%%%%%%%%%%%%%%%%%%%%%%%%%%%%%%%%%%%%%
\begin{table}
\begin{center}
\begin{tabular}{|c|c|}
\hline
$|V_{ud}| = 0.97425\pm 0.00022$ & ${\cal{B}}(B\to X_s \ell^+ \ell^-)_{\rm low} = (1.60 \pm 0.48)\times 10^{-6}$\cite{Lees:2013nxa}\\
$|V_{us}| = 0.2252\pm 0.0009$ & ${\cal{B}}(B\to X_s \ell^+ \ell^-)_{\rm high} = (0.57 \pm 0.16)\times 10^{-6}$\cite{Lees:2013nxa}\\
$|V_{cd}| = 0.230\pm 0.011$ &$10^{9}\, {\rm GeV^2} \times \langle  \frac{d{\cal{B}}}{dq^2} \rangle(B\to K \mu^+ \mu^-)_{\rm low} = 18.7 \pm 3.6$\cite{Aaij:2014pli}\\
$|V_{cs}| = 1.006\pm 0.023$ &$10^{9}\, {\rm GeV^2} \times \langle  \frac{d{\cal{B}}}{dq^2} \rangle(B\to K \mu^+ \mu^-)_{\rm high} = 9.5 \pm 1.7$\cite{Aaij:2014pli}\\
$|V_{ub}| = 0.00382\pm 0.00021$ & ${\cal{B}}(B^+\to \pi^+ \mu^+ \mu^-) = (2.60 \pm 0.61)\times 10^{-8}$ \cite{LHCb:2012de}\\
$|V_{cb}| = (40.9\pm 1.0)\times 10^{-3}$  & ${\cal{B}}(K^+\to \pi^+\nu\bar\nu) = (1.7 \pm 1.1)\times 10^{-10}$ \\
$\gamma=(68.0 \pm 11.0)^{\circ}$ &  ${\cal{B}}(K_L\to \mu^+ \mu^-) = (0 \pm 1.56)\times 10^{-9}$ \cite{Isidori:2003ts}\\
$|\epsilon_k|\times 10^{3} = 2.228 \pm 0.011$ & ${\cal{B}}(B_s\to \mu^+ \mu^-) = (2.9\pm0.7)\times 10^{-9}$ \cite{Aaij:2013aka,Chatrchyan:2013bka,CMS:2014xfa} \\
$\Delta{M_d} = (0.507 \pm 0.004)\, {\rm ps}^{-1}$\cite{Amhis:2012bh} &  ${\cal{B}}(B_d\to \mu^+ \mu^-) = (3.9\pm1.6)\times 10^{-10}$ \cite{Aaij:2013aka,Chatrchyan:2013bka,CMS:2014xfa} \\
$\Delta{M_s} = (17.72 \pm 0.04)\, {\rm ps}^{-1}$\cite{Amhis:2012bh} & ${\cal{B}}(B \to X_s\,\gamma) = (3.55 \pm 0.26)\times 10^{-4}$\\
$S_{J/\psi\,\phi}= 0.00 \pm 0.07$\cite{Amhis:2012bh}& ${\cal{B}}(B\to \tau \,\bar{\nu}) = (1.14 \pm 0.22)\times 10^{-4}$ \cite{Amhis:2012bh}\\
$S_{J/\psi\,K_S}= 0.68 \pm 0.02$\cite{Amhis:2012bh}&  $A^b_{sl}=(-4.96 \pm 1.69)\times 10^{-3}$ \cite{Abazov:2013uma}\\
$S= 0.00 \pm 0.11$ & $A_b = 0.923 \pm 0.020$\cite{Abbaneo:2001ix}\\
$T= 0.02 \pm 0.12$ & $R_b = 0.2164 \pm 0.0007$\cite{Abbaneo:2001ix}\\
\hline 
\end{tabular} 
\caption{Experimental values of flavor-physics observables used as
  constraints.  For $V_{ub}$ we use the weighted average from the
  inclusive and exclusive semileptonic decays, $V_{ub}^{inc}=(44.1\pm
  3.1) \times 10^{-4}$ and $V_{ub}^{exc}=(32.3 \pm 3.1)\times
  10^{-4}$. When not explicitly stated, we take the inputs from the
  Particle Data Group \cite{pdg}. Wherever there are asymmetric
  experimental errors, they are symmetrized by taking the largest side
  error. Also, wherever there is more than one source of uncertainty,
  the total error is obtained by adding these in quadrature.}
\label{tab1}
\end{center}
\end{table}
%%%%%%%%%%%%%%%%%%%%%%%%%%%%%%%%%%%%%%%%%%%%%%%%%%%%%%

%%%%%%%%%%%%%%%%
\begin{table}
\begin{center}
\begin{tabular}{|c|c|c|}
\hline
$q^2 = 0.1$-2 GeV$^2$ & $q^2 = 2$-4.3 GeV$^2$ & $q^2 = 4.3$-8.68 GeV$^2$  \\
\hline 
 $\langle\frac{d{\cal{B}}}{dq^2}\rangle$ = $(0.60 \pm 0.10)\times 10^{-7}$
& $\langle\frac{d{\cal{B}}}{dq^2}\rangle$ = $(0.30 \pm 0.05)\times 10^{-7}$ &$\langle\frac{d{\cal{B}}}{dq^2}\rangle$ = $(0.49 \pm 0.08)\times 10^{-7}$\\
$\langle F_{L} \rangle$ = $ 0.37 \pm 0.11$& $\langle F_{L} \rangle$ = $ 0.74 \pm 0.10$ & 
$\langle F_{L} \rangle$ = $ 0.57 \pm 0.08 $ \\
$\langle P_1 \rangle$ = $ -0.19 \pm 0.40 $ & $\langle P_1 \rangle$ = $ -0.29 \pm 0.65$  & 
$\langle P_1 \rangle$ = $ 0.36 \pm 0.31$ \\
$\langle P_2 \rangle$ = $ 0.03 \pm 0.15$ & $\langle P_2 \rangle$ = $ 0.50 \pm 0.08$ & 
$\langle P_2 \rangle$ = $ -0.25 \pm 0.08$ \\
$\langle P_4' \rangle$ = $0.00 \pm 0.52$ & $\langle P_4' \rangle$ = $0.74 \pm 0.60$ & 
$\langle P_4' \rangle$ = $1.18 \pm 0.32$ \\ 
$\langle P_5' \rangle$ = $0.45 \pm 0.24$ & $\langle P_5' \rangle$ = $0.29 \pm 0.40$ &
$\langle P_5' \rangle$ = $-0.19 \pm 0.16$ \\
$\langle P_6' \rangle$ = $0.24 \pm 0.23$ & $\langle P_6' \rangle$ = $-0.15 \pm 0.38 $ & 
$\langle P_6' \rangle$ = $0.04 \pm 0.16$ \\
$\langle P_8' \rangle$ = $-0.12 \pm 0.56$ & $\langle P_8' \rangle$ = $-0.3 \pm 0.60$ &
$\langle P_8' \rangle$ = $0.58 \pm 0.38$ \\
\hline
$q^2 = 14.18$-16 GeV$^2$ & $q^2 = 16$-19 GeV$^2$&  \\
\hline
 $\langle\frac{d{\cal{B}}}{dq^2}\rangle$ = $(0.56 \pm 0.10)\times 10^{-7}$
& $\langle\frac{d{\cal{B}}}{dq^2}\rangle$ = $(0.41 \pm 0.07)\times 10^{-7}$ &\\
 $\langle F_{L} \rangle$ = $ 0.33 \pm 0.09$
& $ \langle F_{L} \rangle$ = $ 0.38 \pm 0.09$ &\\
 $\langle P_1 \rangle$ = $ 0.07 \pm 0.28$
& $ \langle P_1 \rangle$ = $ -0.71 \pm 0.36$ &\\
 $\langle P_2 \rangle$ = $ -0.50 \pm 0.03$
& $ \langle P_2 \rangle$ = $ -0.32 \pm 0.08$ &\\
$\langle P_4' \rangle$ = $-0.18 \pm 0.70$
& $ \langle P_4' \rangle$ = $0.70 \pm 0.52$ &\\
 $\langle P_5' \rangle$ = $-0.79 \pm 0.27$
& $ \langle P_5' \rangle$ = $-0.60 \pm 0.21 $ &\\
$\langle P_6' \rangle$ = $0.18 \pm 0.25$
& $ \langle P_6' \rangle$ = $-0.31 \pm 0.39$ &\\
 $\langle P_8' \rangle$ = $-0.40 \pm 0.60$ 
& $ \langle P_8' \rangle$ = $0.12 \pm 0.54$ &\\
 \hline
\end{tabular}
\caption{Experimental values of the observables in $B \to K^*\, \mu^+
  \,\mu^-$ used as constraints. These are taken from
  Refs.~\cite{Aaij:2013iag,Aaij:2013qta}. Here the errors have been
  symmetrized by taking the largest side error. Also, wherever there
  is more than one source of uncertainty, the total error is obtained
  by adding these in quadrature.}
\label{bkstar}
\end{center}
\end{table}

%%%%%%%%%%%%%%%%%%%%%%%%%%%%%%%%%%%%%%%%%%%%%%%%%%%%%%%%%%%%%%%%%%%%%%%%%%%%%%%%%%%%%%%%%%%%%%%%
\begin{table}
\begin{center}
\begin{tabular}{|c|c|}
\hline
$G_F = 1.16637 \times 10^{-5}$ Gev$^{-2}$&  $\tau_{B_s}=(1.497\pm 0.026)$ ps  \\
$ \sin^2\theta_w = 0.23116$              &  $\tau_{B^{\pm}}=(1.641\pm0.008)$ ps\\
$ \alpha(M_Z) = \frac{1}{127.9}$         &  $ \eta_t = 0.5765$ \cite{Buras:1990fn}  \\
$ \alpha_s(M_Z) = 0.1184$                &  $ \eta_{ct} = 0.496 \pm 0.047$ \cite{Brod:2010mj}\\
$m_t(m_t) = 163 $ GeV                    &  $f_K = 0.1561 \pm 0.0011$ \cite{Laiho:2009eu}\\
$m_c(m_c) = 1.275 \pm 0.025 $ GeV                     &  $B_K = 0.767 \pm 0.010$ \cite{Laiho:2009eu}\\
$m_b(m_b) = 4.18 \pm 0.03 $ GeV                  &  $\Delta M_K = (0.5292 \pm 0.0009)\times 10^{-2} \, {\rm ps}^{-1}$\\
$M_W = 80.385$ GeV                       &  $ \kappa_{\epsilon}=0.94 \pm 0.02$ \cite{Buras:2008nn,Buras:2010pza}\\
$M_Z = 91.1876$ GeV                      &  $ \kappa_+ = (5.36\pm0.026)\times 10^{-11}$ \cite{Mescia:2007kn}\\
$M_K = 0.497614$ GeV                     &  $ \kappa_{\mu} = (2.009\pm 0.017) \times 10^{-9}$ \cite{Gorbahn:2006bm}\\
$M_{K^*} = 0.89594$ GeV                  &  $f_{bd}=(190.5 \pm 4.2)$ MeV \cite{Aoki:2013ldr}\\
$M_D = 1.86486$ GeV                      &  $f_{bs}=(227.7\pm4.5)$ MeV \cite{Aoki:2013ldr}\\
$M_{B_d} = 5.27917$ GeV                  &  $f_{\bd}\sqrt{B_{\bd}} = (0.216 \pm 0.015)$ {\rm GeV}\cite{Aoki:2013ldr}\\
$M_{B_s} = 5.36677$ GeV                  &  $ \xi=1.268\pm 0.063$ \cite{Aoki:2013ldr}\\
$M_{B^{\pm}} = 5.27926$ GeV              &  ${\cal{B}}(B\to X_c \ell \nu) = (10.61 \pm 0.17)\times 10^{-2}$\\
$m_{\mu} = 0.105$ GeV                    &  $m_c/m_b=0.29 \pm 0.02$\\
$m_{\tau} = 1.77682$ GeV                 & \\
$\tau_{B_d}= (1.519\pm 0.007)$ ps                   & \\
\hline
\end{tabular}
\caption{Decay constants, bag parameters, QCD corrections and other
  parameters used in our analysis.  When not explicitly stated, we take the
  inputs from the Particle Data Group \cite{pdg}.}
\label{tab3}
\end{center}
\end{table}
%%%%%%%%%%%%%%%%%%%%%%%%%%%%%%%%%%%%%%%%%%%%%%%%%%%%

%%%%%%%%%%%%%%%%%%%%%%%%%%%%%%%%%%%%%%%%%%%%%%%%%%%%%%%%%%%%%%%%%%%%%%%%%%
\subsection{\bf Direct measurements of the CKM elements }
%%%%%%%%%%%%%%%%%%%%%%%%%%%%%%%%%%%%%%%%%%%%%%%%%%%%%%%%%%%%%%%%%%%%%%%%%%

The latest values for the direct measurements of the magnitudes of the
CKM matrix elements can be found in Ref.~\cite{pdg}.  The contribution
to $ \chi^2_{\rm total}$ from these measurements is given by
\bea
\chi^2_{\rm CKM} &=& \Big( \frac{|V_{us}|-0.2252}{0.0009} \Big)^2 
+ \Big( \frac{|V_{ud}|-0.97425}{0.00022} \Big)^2
+ \Big( \frac{|V_{cs}|-1.006}{0.023} \Big)^2 
\nonumber\\&&
+~\Big( \frac{|V_{cd}|-0.230}{0.011} \Big)^2
+ \Big( \frac{|V_{ub}|-0.00382}{0.00021} \Big)^2 
+ \Big( \frac{|V_{cb}|-0.0409}{0.001} \Big)^2 \;.
\eea

%%%%%%%%%%%%%%%%%%%%%%%%%%%%%%%%%%%%%%%%%%%%%%%%%%%%%%%%%%%%%%%%%%%%
\subsection{\bf \boldmath CP violation in $K_L \to \pi\pi$: $ \epsilon_K$}
%%%%%%%%%%%%%%%%%%%%%%%%%%%%%%%%%%%%%%%%%%%%%%%%%%%%%%%%%%%%%%%%%%%

In the VuQ model, the mixing amplitude $M^{12}_K$ is modified due to
an additional contribution coming from a virtual $t'$ quark in the box
diagram. There is a sizeable LD contribution to the mass difference
$\Delta M_K$ in the $K$ system, for which, at present, there is no
definitive estimate. We therefore do not include $\Delta M_K$ in our
analysis. However, $|\epsilon_K|$, the parameter describing the
mixing-induced $CP$ asymmetry in neutral $K$ decays, and which is
proportional to ${\rm Im}(M^{12}_K)$, is theoretically clean and is a
well-measured quantity. The theoretical expression for $|\epsilon_K|$
in the presence of a $t'$ quark is given in Refs.~\cite{SM4,SAGMN2}.

To calculate the contribution of $|\epsilon_K|$ to $ \chi^2_{\rm
  total}$, we use the quantity
\beq
K_{\rm mix} = \frac{12\sqrt{2}\pi^2(\Delta M_K)_{\rm exp} |\epsilon_K|}
{G^2_F M^2_W f^2_K M_K \hat{B}_K k_\epsilon}
- {\rm Im}\left[\eta_c (V_{cs}V^*_{cd})^2 S(x_c)\right] ~.
\eeq
With the experimental and theoretical inputs given in Tables
\ref{tab1} and \ref{tab3}, we find
\beq
K_{\rm mix, \, exp} = (1.69\pm0.05) \times 10^{-7} \; .
\eeq
The QCD correction $ \eta_{ct}$ appears in the theoretical expression
of $|\epsilon_K|$. In order to take its error into account, we
consider it to be a parameter and have added a contribution to
$\chi^2_{\rm total}$.  We hold the other QCD correction $ \eta_t$
fixed to its central value because its error is very small. The total
contribution to $ \chi^2_{\rm total}$ from $|\epsilon_K|$ is then
\beq
\chi^2_{|\epsilon_K|} = \Big( \frac{K_{\rm mix} - 1.69 \times 10^{-7}}
{0.05 \times 10^{-7}} \Big)^2 + \Big( \frac{\eta_{ct} - 0.496}{0.047} \Big)^2 ~.
\eeq

%%%%%%%%%%%%%%%%%%%%%%%%%%%%%%%%%%%%%%%%%%%%%%%%%%%%%%%%%%%%%%%%%%%%%%%%%%
\subsection{\bf \boldmath Branching fraction of the decay $K^+ \to \pi^+ \nu {\bar \nu}$}
%%%%%%%%%%%%%%%%%%%%%%%%%%%%%%%%%%%%%%%%%%%%%%%%%%%%%%%%%%%%%%%%%%%%%%%%%%

In Refs.~\cite{Rein:1989tr,Hagelin:1989wt}, it was shown that the LD
contribution to ${\cal B}(K^+\to \pi^+\nu\bar{\nu})$ is suppressed --
it is three orders of magnitude smaller than the short-distance (SD)
contribution. The SM prediction for this observable is therefore under
good control.  The decay $K^+ \to \pi^+ \nu {\bar \nu}$ occurs via
loops containing virtual heavy particles, and hence is sensitive to
the $t'$ quark.  The theoretical expression for ${\cal B}(K^+\to
\pi^+\nu\bar{\nu})$ in the presence of a $t'$ quark is given in
Refs.~\cite{SM4,SAGMN2}.

With the inputs given in Tables \ref{tab1} and \ref{tab3}, we estimate
\beq
\frac{{\cal B}(K^+\to \pi^+\nu\bar{\nu})}{\kappa_+} = 3.17 \pm 2.05 ~,
\eeq 
where
\beq
\kappa_+ = r_{K^+} \frac{3 \alpha^2 {\cal B}(K^+\to \pi^0 e^+ \nu)}{2 \pi^2 \sin^4 \theta_W} \lambda^8 ~.
\eeq
Here $r_{K^+}=0.901$ encapsulates the isospin-breaking corrections in
relating the branching ratio of $K^+\to \pi^+\nu\bar{\nu}$ to that of
the well-measured decay $K^+\to \pi^0 e^+ \nu$.
 
In order to include ${\cal B}(K^+ \to \pi^+ \nu {\bar \nu})$ in the
fit, we define
\beq
\chi^2_{K^+\to \pi^+\nu \bar{\nu}} =\Big( \frac{[{\cal B}(K^+\to \pi^+\nu\bar{\nu})/\kappa_+] - 3.17}{2.05} \Big)^2\;.
\eeq

%%%%%%%%%%%%%%%%%%%%%%%%%%%%%%%%%%%%%%%%%%%%%%%%%%%%%%%%%%%%%%%%%%%%%%%%%%
\subsection{\bf \boldmath Branching fraction of the decay $K_L \to \mu^+ \mu^-$}
%%%%%%%%%%%%%%%%%%%%%%%%%%%%%%%%%%%%%%%%%%%%%%%%%%%%%%%%%%%%%%%%%%%%%%%%%%

Unlike $K^+\to \pi^+\nu\bar{\nu}$, the decay $K_L \to \mu^+ \mu^-$ is
not cleanly dominated by the SD contribution.  However, it is possible
to estimate the LD contribution to this decay. The absorptive LD
contribution is estimated using $K_L \to \gamma \gamma$, while the
dispersive LD contribution is estimated using chiral perturbation
theory along with the experimental inputs on various $K$ decays.  Due
to uncertainties involved in the extraction of the dispersive
contribution, one can only obtain a conservative upper limit on the SD
contribution to ${\cal B}(K_L \to \mu^+ \mu^-)$, which is $\leq 2.5
\times 10^{-9}$ \cite{Isidori:2003ts}. With all the inputs given in
Tables \ref{tab1} and \ref{tab3}, we estimate
\beq
\frac{{\cal B}(K_L \to \mu^+ \mu^-)}{\kappa_{\mu}} = 0 \pm 0.778 ~, 
\eeq 
where
\beq
\kappa_\mu = \frac{\alpha^2 {\cal B}(K^+ \to \mu^+ \nu_{\mu}) }{\pi^2 \sin^4 \theta_W} 
\frac{\tau(K_L)}{\tau(K^+)} \lambda^8\,.
\eeq

In the VuQ model, the theoretical expression for ${\cal B}(K_L \to
\mu^+ \mu^-)/\kappa_{\mu}$ is given by
\bea
\frac{{\cal B}(K_L \to \mu^+ \mu^-)}{\kappa_{\mu}} &=& 
\Bigg( \frac{{\rm Re}(V_{cd}V^*_{cs})}{\lambda} P_c + \frac{{\rm Re}(V_{td}V^*_{ts})}{\lambda^5}Y(x_t) \nn\\ 
&& \hskip3truecm
+~\frac{{\rm Re}(V_{t'd}V^*_{t's})}{\lambda^5}Y(x_{t'}) \Bigg)^2 ~.
\eea
Here $Y(x)$ is the structure function in the $t$ or $t'$ sector
\cite{Buchalla:1998ba,Inami:1980fz}, while $P_c$ is the corresponding
structure function in the charm sector. Its NNLO QCD-corrected value
is $P_c=0.115\pm 0.018$ \cite{Gorbahn:2006bm}.  In order to include
${\cal B}(K_L \to \mu^+ \mu^-)$ in the fit, we define
\beq
\chi^2_{K_L \to \mu^+ \mu^-} =\Big( \frac{{\cal B}(K_L \to \mu^+ \mu^-)/\kappa_{\mu} - 0}{0.778} \Big)^2 +
\Big( \frac{P_c - 0.115}{0.018} \Big)^2 ~.
\eeq
Thus, the error on $P_c$ has been taken into account by considering it
to be a parameter and adding a contribution to $\chi^2_{\rm total}$.

%%%%%%%%%%%%%%%%%%%%%%%%%%%%%%%%%%%%%%%%%%%%%%%%%%%%%%%%%%%%%%%
\subsection{\bf \boldmath $Z \to b\bar{b}$ decay} 
%%%%%%%%%%%%%%%%%%%%%%%%%%%%%%%%%%%%%%%%%%%%%%%%%%%%%%%%%%%%%%%

Here we include constraints from $R_b$ and ${A_b}$, respectively the
vertex correction and forward-backward asymmetry in $Z \to b \bar{b}$.
The theoretical expressions for $R_{b}$ and $A_{b}$ in the VuQ model
are given in Ref.~\cite{AgSaav}. We have
\beq
\chi^2_{Z \to b b} = \Big( \frac{R_{b}-0.216}{0.001} \Big)^2 + \Big( \frac{A_{b}-0.923}{0.020} \Big)^2.
\eeq

%%%%%%%%%%%%%%%%%%%%%%%%%%%%%%%%%%%%%%%%%%%%%%%%%%%%%%%%%%%%%%%%%%
\subsection{\bf \boldmath $B^0_q$-$\bar B^0_q$ mixing ($q = d,s$)}
%%%%%%%%%%%%%%%%%%%%%%%%%%%%%%%%%%%%%%%%%%%%%%%%%%%%%%%%%%%%%%%%

The theoretical expressions for $M^q_{12}$ ($q = d,s$) in the presence
of a $t'$ quark, which then lead to $\Delta M_d$ and $\Delta M_s$, are
given in Refs.~\cite{SM4,SAGMN2}.  To calculate $\chi^2_{\bd}$ for
$\bd$-$\bdbar$ mixing, we use the quantity
\beq
B^d_{\rm mix} = \frac {6\pi^2 \Delta M_d}{G^2_F M_W^2 M_{B_d} 
\hat{B}_{bd} f_{\bd}^2} \; .
\eeq
With the inputs given in Table~\ref{tab1}, we get
\beq
B^d_{\rm mix,{\rm exp}} =  (9.12249 \pm 1.26905 ) \times 10^{-5} \; ,
\eeq 
leading to
\beq
\chi^2_{\bd} = \Big( \frac{B^d_{\rm mix}-9.12249 \times 10^{-5}}{1.26905 \times 10^{-5}} \Big)^2\;.
\eeq

To take $\bs$-$\bsbar$ mixing into account, we define
\beq
M_R = \frac{\Delta M_s}{\Delta M_d} \frac{M_{B_d}}{M_{B_s}} \frac{1}{\xi^2} ~,
\eeq
whose measured value is
\beq
M_{R,\rm exp}= 21.3831 \pm 2.1321\,.
\eeq
Then
\beq
\chi^2_{M_R} = \Big( \frac{M_R-21.3831}{2.1321} \Big)^2 ~.
\eeq

%%%%%%%%%%%%%%%%%%%%%%%%%%%%%%%%%%%%%%%%%%%%%%%%%%%%%%%%%%%%%%%%%%
\subsection{\bf \boldmath Indirect CP violation in $ \bd \to J/\psi\, K_S$ and $ \bs \to J/\psi\, \phi$}
%%%%%%%%%%%%%%%%%%%%%%%%%%%%%%%%%%%%%%%%%%%%%%%%%%%%%%%%%%%%%%%%

The theoretical expressions for $M^q_{12}$ ($q = d,s$) in the VuQ
model are discussed in the previous subsection. In the SM, indirect CP
violation in $ \bd \to J/\psi\, K_S$ and $ \bs \to J/\psi\, \phi$
probes $\sin 2\beta$ and $\sin 2\beta_s$, respectively. With NP, we
have
\beq
S_{J/\psi\, K_S} = \frac{{\rm Im}(M^d_{12})}{M^d_{12}}, \hskip 30pt
S_{J/\psi\, \phi} = -\frac{{\rm Im}(M^s_{12})}{M^s_{12}}\;.
\eeq
The experimentally-measured values of $\sin 2\beta$ and $\sin
2\beta_s$ are given in Ref.~\cite{pdg}. Then
\beq
\chi^2_{\sin 2\beta} = \Big( \frac{S_{J/\psi\, K_S}-0.68}{0.02}\Big)^2, \hskip 30pt
\chi^2_{\sin 2\beta_s} = \Big( \frac{S_{J/\psi\, \phi}-0.00}{0.07} \Big)^2 ~.
\eeq

%%%%%%%%%%%%%%%%%%%%%%%%%%%%%%%%%%%%%%%%%%%%%%%%%%%%%%%%%%%%%%%%%%%%
\subsection{\bf \boldmath CKM angle $\gamma$}
%%%%%%%%%%%%%%%%%%%%%%%%%%%%%%%%%%%%%%%%%%%%%%%%%%%%%%%%%%%%%%%%%%%

In the Wolfenstein parametrization, the CKM angle $\gamma =
\tan^{-1}(\eta/\rho)$, which is the argument of $V_{ub}$.  As this
angle is measured in tree-level decays, its value is unchanged with
the addition of a vector isosinglet up-type quark.  Therefore the
$\chi^2$ of $\gamma$ is given by
\beq
\chi^2_{\gamma} = \Big( \frac{\delta_{ub}- 68~(\pi/180)}{11~(\pi/180)} \Big)^2\;.
\eeq

%%%%%%%%%%%%%%%%%%%%%%%%%%%%%%%%%%%%%%%%%%%%%%%%%%%%%%%%%%%%%%%%%%%%%%%%%%
\subsection{\bf \boldmath Branching ratio of $B \to X_s\, l^+ \,l^-$ ($l=e,\mu$)}
%%%%%%%%%%%%%%%%%%%%%%%%%%%%%%%%%%%%%%%%%%%%%%%%%%%%%%%%%%%%%%%%%%%%%%%%%%

The quark-level transition $b \to s\,l^+ \, l^-$ can occur only at
loop level within the SM, so that it can be used to test higher-order
corrections to the SM, and to constrain various NP models.  Within the
SM, the effective Hamiltonian for this transition can be written as
\beq
{\cal H}_{eff} =  - \frac{4 G_F}{\sqrt{2}} V_{ts}V^*_{tb}
\sum_{i=1}^{10} C_i(\mu) \,  O_i(\mu)\;,
\label{Heffbs}
\eeq
where the form of the operators $O_i$ and the expressions for
calculating the coefficients $C_i$ are given in
Ref.~\cite{Buras:1994dj}. In the VuQ model only the values of the
Wilson coefficients $C_{7,8,9,10}$ are changed via the virtual
exchange of the $t'$ quark. The modified Wilson coefficients in the
vector-singlet up-quark model can then be written as \cite{SM4,SAGMN2}
\beq 
C^{\rm tot}_{j}(\mu_b) = C_j(\mu_b) +
\frac{V_{t^{'}s}V_{t^{'}b}^{*}}{V_{ts}V_{tb}^{*}} C_j^{t'}(\mu_b)\,,
\label{ctot}
\eeq 
where $j=7,8,9,10$. The new wilson coefficients $C_j^{t'}$ can be
calculated from the expression of $C_j$ by replacing $m_t$ by
$m_{t'}$.  
       
The inclusive decay mode $B \to X_s\, l^+ \,l^-$ has relatively small
theoretical errors as compared to the exclusive decay modes $B \to
(K,K^*)\, l^+ \,l^-$.  However, the inclusive decays are less readily
accessible experimentally.  The branching ratio of $B \to X_s\, l^+
\,l^-$ has been measured by the Belle and BaBar Collaborations using
the sum-of-exclusive technique. The latest Belle measurement uses only
~25\% of its final data set \cite{Iwasaki:2005sy}. The BaBar
Collaboration has recently published the measurement of ${\cal B}(B
\to X_s \, l^+ \,l^-)$ using the full data set, which corresponds
to $471 \times 10^{6}$ $B\bar{B}$ events \cite{Lees:2013nxa}.  This is
an update of their previous result, which was based on a data sample
of $89 \times 10^{6}$ $B\bar{B}$ events \cite{Aubert:2004it}.

The prediction for the branching ratio is relatively cleaner in the
low-$q^2$ (1 $ \rm GeV^2$ $ \leq$ $q^2$ $ \leq$ 6 $ \rm GeV^2$) and
high-$q^2$ (14.2 $ \rm GeV^2$ $ \leq$ $q^2$ $ \leq$ $m_b^2$)
regions. We consider both regions in the fit.  The theoretical
predictions for ${\cal B}(B \to X_s \, l^+ \,l^-)$ are computed using
the program {\bf SuperIso} \cite{Mahmoudi:2007vz,Mahmoudi:2008tp}, in
which the higher-order and power corrections are implemented following
Refs.~\cite{Ghinculov:2003qd, Huber:2005ig}, while the electromagnetic
logarithmically-enhanced corrections are taken from
Refs.~\cite{Huber:2007vv}. Bremsstrahlung contributions are
implemented following Refs.~\cite{Asatryan:2002iy}.  

The contribution to $\chi^2_{\rm total}$ is
\bea
\chi^2_{B \to X_s\, l^+ \,l^-} & = & \Big( \frac{{\cal B}(B  \to X_s \, l^+ \,l^-)_{\rm low}-1.6\times 10^{-6}}{0.49\times 10^{-6}} \Big)^2 \nn\\
&& \hskip2truecm 
+~\Big( \frac{{\cal B}(B  \to X_s \, l^+ \,l^-)_{\rm high}-0.57\times 10^{-6}}{0.23\times 10^{-6}} \Big)^2 ~,
\eea
where we have added a theoretical error of $7\%$ to ${\cal B}(B \to
X_s\,l^+ \,l^-)_{\rm low}$, which includes corrections due to the
renormalization scale and quark masses, and a theoretical error of
$30\%$ to ${\cal B}(B \to X_s\,l^+ \,l^-)_{\rm high}$, which includes
the non-perturbative QCD corrections.

%%%%%%%%%%%%%%%%%%%%%%%%%%%%%%%%%%%%%%%%%%%%%%%%%%%%%%%%%%%%%%%%%%%%%%%%%%
\subsection{\bf \boldmath Branching ratio of $B \to X_s\, \gamma$ }
%%%%%%%%%%%%%%%%%%%%%%%%%%%%%%%%%%%%%%%%%%%%%%%%%%%%%%%%%%%%%%%%%%%%%%%%%%

The quantity we use for $B  \to X_s \, \gamma$ is 
\beq
{\tilde R} =\frac{\pi\,f(\hat{m_c})\,\kappa(\hat{m}_c)}{6\,\alpha}
\frac{{\cal B}(B \to X_s \gamma)}{{\cal B}(B \to X_c e \bar \nu_e)}\;,
\eeq
where the ratio of the two branching fractions is taken in order to
reduce the large uncertainties arising from $b$-quark mass.  The
theoretical expression for ${\cal B}(B \to X_s \, \gamma)$ 
is given in Refs.~\cite{SM4,SAGMN2}. From this, one can
deduce the expression for ${\tilde R}$.  The measured value of ${\tilde R}$ is
\beq
{\tilde R}_{\rm exp} =  0.1069 \pm 0.0120 \;,
\eeq
where we have added an overall correction of 5\% due to the
non-perturbative terms.  The contribution to $\chi^2_{\rm total}$ is
\beq
\chi^2_{B \to X_s \gamma}  =\Big( \frac{{\tilde R} - 0.1069}
{0.0120} \Big)^2\; .
\eeq

%%%%%%%%%%%%%%%%%%%%%%%%%%%%%%%%%%%%%%%%%%%%%%%%%%%%%%%%%%%%%%%%%%%%%%%%%%
\subsection{\bf \boldmath Branching ratio of $B \to K\, \mu^+ \,\mu^-$}
%%%%%%%%%%%%%%%%%%%%%%%%%%%%%%%%%%%%%%%%%%%%%%%%%%%%%%%%%%%%%%%%%%%%%%%%%%

The theoretical expression for $\langle d{\cal B}/dq^2\rangle(B \to K
\, \mu^+ \,\mu^-)$ in the SM is given in
Refs.~\cite{Bobeth:2007dw,Bobeth:2011nj}, and can be adapted
straightforwardly to the VuQ model.  The predictions for the branching
ratio are relatively cleaner in the low-$q^2$ (1.1 $ \rm GeV^2$ $
\leq$ $q^2$ $ \leq$ 6 $ \rm GeV^2$) and the high-$q^2$ (15 $ \rm
GeV^2$ $ \leq$ $q^2$ $ \leq$ 22 $ \rm GeV^2$) regions. Here, we
consider both regions in the fit. We use the recent LHCb measurements
of $\langle d{\cal B}/dq^2\rangle(B \to K \, \mu^+ \,\mu^-)$
\cite{Aaij:2014pli}.

Our analysis of $B \to K\, \mu^+ \,\mu^-$ in the low-$q^2$ region is
based on QCD factorization (QCDf) \cite{Beneke:2001at}. The
factorizable and non-factorizable corrections of $O(\alpha_s)$ are
included in our numerical analysis following
Refs.~\cite{Bobeth:2007dw,Beneke:2001at}. In the high-$q^2$ region,
following Ref.~\cite{Bobeth:2011nj}, we use the improved Isgur-Wise
relation between the form factors which are determined using
light-cone QCD sum-rule calculations extrapolated to the high-$q^2$
region.  The contribution to $ \chi^2_{\rm total}$ from $B \to K \,
\mu^+ \,\mu^-$ is
\bea
\chi^2_{B \to K\, \mu^+ \,\mu^-} & = & \Big( \frac{\langle \frac{d{\cal B}}{dq^2}\rangle(B \to K \, \mu^+ \,\mu^-)_{\rm low}-18.7\times 10^{-9}}{6.67\times 10^{-9}} \Big)^2 \nn\\
&& \hskip2truecm 
+~\Big( \frac{\langle \frac{d{\cal B}}{dq^2}\rangle(B \to K \, \mu^+ \,\mu^-)_{\rm high}-9.5\times 10^{-9}}{3.32\times 10^{-9}} \Big)^2 ~,
\eea
where, following Refs.~\cite{Bobeth:2007dw,Bobeth:2011nj}, we have
included a theoretical error of $30\%$ in both low- and high-$q^2$
bins. This is due mainly to uncertainties in the $B \to K$ form
factors.

%%%%%%%%%%%%%%%%%%%%%%%%%%%%%%%%%%%%%%%%%%%%%%%%%%%%%%%%%%%%%%%%%%%%%%%%%%
\subsection{\bf \boldmath Constraints from  $B \to K^*\, \mu^+ \,\mu^-$}
%%%%%%%%%%%%%%%%%%%%%%%%%%%%%%%%%%%%%%%%%%%%%%%%%%%%%%%%%%%%%%%%%%%%%%%%%%

The recent LHCb measurements of new angular observables in $B \to
K^*\, \mu^+ \,\mu^-$ exhibit small tensions with the SM predictions
\cite{Aaij:2013qta,Descotes-Genon:2013wba}.  These tensions can be due
to NP, but can also be attributed to underestimated hadronic power
corrections, or can simply be a statistical fluctuation.  In our
analysis, we include all measured observables in $B \to K^*\, \mu^+
\,\mu^-$ in the low- and high-$q^2$ regions. The experimental results
for $B \to K^*\, \mu^+ \,\mu^-$ decay are given in Table \ref{bkstar},
and are taken from Refs.~\cite{Aaij:2013iag,Aaij:2013qta}.

The complete angular distribution for the decay $B \to K^*\, \mu^+
\,\mu^-$ is described by four independent kinematic variables: the
lepton-pair invariant mass squared $q^2$, two polar angles
$\theta_\mu$ and $\theta_K$, and the angle between the planes of the
dimuon and $K\pi$ decays, $\phi$. The differential decay distribution
of $B \to K^*\, \mu^+ \,\mu^-$ can be written as
\begin{equation}
  \label{eq:differential decay rate}
  \frac{d^4\Gamma[B \to K^{*}(\to K \pi)\mu^+\mu^-]}
       {d q^2\, d\ctl\, d\ctk\, d\phi} =
  \frac{9}{32\pi}  J(q^2 , \theta_l, \theta_K, \phi)\,.
\end{equation}
where the angular-dependent term can be written as 
\bea
 J(q^2 , \theta_l, \theta_K, \phi)&=&J_{1s}\sstk + J_{1c}\cstk + (J_{2s}\sstk + J_{2c}\cstk)\cttl \nonumber \\
&& \hskip-2truecm +~J_3\sstk\sstl\ctp \nonumber + J_4\sttk\sttl\cp \nonumber \\
&& \hskip-2truecm +~J_5\sttk\stl\cp + (J_{6s}\sstk + J_{6c}\cstk)\ctl \\
&& \hskip-2truecm +~J_7\sttk\stl\sp + J_8\sttk\sttl\sp + J_9\sstk\sstl\stp ~. \nn
\eea
For massless leptons, the $J_i$'s depend on the six complex $K^*$ spin
amplitudes $A_{\parallel}^{L,R}, A_{\perp}^{L,R}$ and $A_0^{L,R}$. For
example,
\begin{equation}
 J_{1s} = \frac{3}{4}[|A_{\perp}^{L}|^2+|A_{\parallel}^{L}|^2+|A_{\perp}^{R}|^2+|A_{\parallel}^{R}|^2] ~.
\end{equation}
For massive leptons, the additional amplitude $A_t$ has to be
introduced. In our analysis, the muon mass is included.
  
The analysis of $B \to K^*\,\mu^+ \,\mu^-$ in the low-$q^2$ region is
based on QCDf \cite{Beneke:2001at} and its quantum
field-theoretical formulation, Soft-Collinear Effective Theory
(SCET). In the limits of a heavy $b$ quark and an energetic $K^*$
meson \cite{Charles:1998dr,Charles:1999gy,Dugan:1990de}, the form
factors can be expanded in the small ratios $\lqcd/m_b$ and $\lqcd/E$,
where $E$ is the energy of the $K^*$ meson. At leading order in
$1/m_b$ and $\alpha_s$, the seven {\it a-priori} independent $B\to
K^*$ form factors reduce to two universal form factors $\xi_{\bot,\|}$
\cite{Charles:1998dr,Charles:1999gy,Dugan:1990de,Beneke:2000wa,Beneke:2004dp}. The
symmetry-breaking corrections of $O(\alpha_s)$, both factorizable and
non-factorizable, are included in our numerical analysis following
Ref.~\cite{Beneke:2001at}. Regarding the $\lqcd/m_b$ corrections to
the QCDf amplitudes, we do not have any means to calculate them in
general. These power corrections can only be estimated by combining
QCDf/SCET results with a QCD sum rule approach, see
Refs.~\cite{Egede:2008uy,Egede:2010zc}.

The analysis of $B \to K^*\,\mu^+ \,\mu^-$ in the high-$q^2$ region is
based on the heavy-quark effective theory framework by Grinstein and
Pirjol \cite{Grinstein:2004vb}. It was shown in
Refs.~\cite{Grinstein:2004vb, Beylich:2011aq} that an operator product
expansion is applicable, which allows one to obtain the $B \to K^*
\mu^+\mu^-$ matrix elements in a systematic expansion in $ \alpha_s$
and in $ \lqcd/m_b$.  The leading $ \lqcd/m_b$ corrections are
parametrically suppressed and contribute only at the few percent
level. The improved Isgur-Wise relations between the form factors at
leading order in $ 1/m_b$ lead to simple expressions for the $K^*$
spin amplitudes to leading order in $1/m_b$
\cite{Bobeth:2010wg,Bobeth:2011gi,Bobeth:2012vn}.  For the form
factors in the high-$q^2$ region, we have used the recent lattice
results \cite{Horgan:2013hoa,Horgan:2013pva}.

Of course, these theoretical predictions have errors associated with
them \cite{Egede:2010zc,Bobeth:2010wg,Hurth:2014vma,Hurth:2013ssa,
  Descotes-Genon:2013vna,Hurth:2012jn,DescotesGenon:2012zf}.  The main
sources of uncertainties in the low-$q^2$ region, excluding
uncertainties due to CKM matrix elements, are (i) the form factors,
(ii) the unknown $1/m_b$ subleading corrections, (iii) the quark
masses, and (iv) the renormalization scale $ \mu_b$.  In the
high-$q^2$ region, there is an additional subleading correction of
$O(1/m_b)$ to the improved Isgur-Wise form factor relations.  For each
$B \to K^*\,\mu^+ \,\mu^-$ observable $O_j$, the theoretical error is
incorporated in the fit by multiplying the theoretical result by
$(1\pm X_j)$, where $X_j$ is the total theoretical error corresponding
to the $j^{\rm th}$ observable and can be easily estimated using Table
II of Ref.~\cite{Hurth:2014vma}.

For $B \to K^* \, \mu^+ \,\mu^-$, we use the observables $\langle d{\cal B}/dq^2\rangle$, 
$P_1$, $P_2$, $P'_4$, $P'_5$, $P'_6$, $P'_8$ and $F_L$ in the low-$q^2$ bins 0.1-2 GeV$^2$,
2.0-4.3 GeV$^2$, 4.3-8.68 GeV$^2$, and the high-$q^2$
bins 14.18-16 GeV$^2$ and 16-19 GeV$^2$.  The SM theoretical
expressions for all observables in $B \to K^*\, \mu^+ \,\mu^-$ in the
low and high-$q^2$ regions are given in \cite{Descotes-Genon:2013vna},
and are straightforwardly adapted to the VuQ model by modifying the
values of the Wilson coefficients as in Eq.~(\ref{ctot}). 
The theoretical predictions for all the $B \to K^* \, \mu^+ \,\mu^-$ observables
are computed using the program {\bf SuperIso}
\cite{Mahmoudi:2007vz,Mahmoudi:2008tp}. For each
bin, we compute the flavor observables and define the $\chi^2$ as
\begin{equation}
\chi^2_{B  \to K^* \, \mu^+ \,\mu^-} \displaystyle =
\sum_{\rm bins} \quad \Bigl[\sum_{j \in ({B\to K^* \mu^+ \mu^- \,{\rm obs.}})}\Bigl(\frac {O_j^{\rm exp} - O_j^{\rm th}}{\sigma_i}\Bigr)^2\Bigr]
\end{equation}

%%%%%%%%%%%%%%%%%%%%%%%%%%%%%%%%%%%%%%%%%%%%%%%%%%%%%%%%%%%%%%%%%%%%%%%%%%%%%
\subsection{\bf \boldmath Branching ratio of $B^+ \to \pi^+ \,\mu^+ \, \mu^-$ }
%%%%%%%%%%%%%%%%%%%%%%%%%%%%%%%%%%%%%%%%%%%%%%%%%%%%%%%%%%%%%%%%%%%%%%%%%%

The quark-level transition $b \to d \mu^+\, \mu^-$ gives rise to the
inclusive semi-leptonic decay ${B}_d^0 \to X_d \,\mu^+ \, \mu^-$, to
exclusive semi-leptonic decays such as ${ B}_d^0 \to \pi^0\,\mu^+
\,\mu^-$, and also to the purely leptonic decay ${ B}_d^0 \to \mu^+ \,
\mu^-$.  However, so far, none of these decays have been observed. We
only have an upper bound on their branching ratios
\cite{Wei:2008nv,Lees:2013lvs}.  Recently, LHCb has observed the $B^+
\to \pi^+\, \mu^+\, \mu^-$ decay with measured branching ratio of
$(2.3 \pm 0.6 \pm 0.1) \times 10^{-8}$ \cite{LHCb:2012de}.  This is
the first measurement of any decay channel induced by $b \to d
\,\mu^+\, \mu^-$.

The effective Hamiltonian for the process $b \to d\, \mu^+\, \mu^-$
and the modified Wilson coefficients in the VuQ model can be
respectively obtained from Eqs.~(\ref{Heffbs}) and (\ref{ctot}) by
replacing $s$ by $d$. The theoretical expression for 
${\cal B}(B^+ \to \pi^+\, \mu^+\, \mu^-)$ is given in Ref.~\cite{Wang:2007sp}.
The contribution to $\chi^2_{\rm total}$ is
\beq
\chi^2_{B^+ \to \pi^+\, \mu^+\, \mu^-}  =\Big( \frac{{\cal B}(B^+ \to \pi^+\, \mu^+\, \mu^-) - 2.3\times 10^{-8}}
{0.66\times 10^{-8}} \Big)^2\; ,
\eeq
where, following Ref.~\cite{Wang:2007sp}, we have included a
theoretical error of $10\%$ in ${\cal B}(B^+ \to \pi^+\, \mu^+\,
\mu^-)$. This is due to uncertainties in the $B^+ \to \pi^+$ form
factors \cite{Ball:2004ye}.

%%%%%%%%%%%%%%%%%%%%%%%%%%%%%%%%%%%%%%%%%%%%%%%%%%%%%%%%%%%%%%%%%%%%%%%%%%%%%
\subsection{\bf \boldmath Branching ratio of $B_q \to \mu^+ \,\mu^-$ $(q=s,d)$ }
%%%%%%%%%%%%%%%%%%%%%%%%%%%%%%%%%%%%%%%%%%%%%%%%%%%%%%%%%%%%%%%%%%%%%%%%%%

The branching ratio of $B_q \to \mu^+ \,\mu^-$ in the VuQ model is
given by
\beq
{\cal B}(B_q \to \mu^+ \,\mu^-) = \frac{G^2_F \alpha^2 M_{B_q} m_\mu^2 f_{bq}^2 \tau_{B_q}}{16 \pi^3} 
|V_{tq}V^*_{tb}|^2  \sqrt{1 - 4 (m_\mu^2/M_{B_q}^2)} 
|C^{\rm tot, q}_{10}|^2 ~,
\eeq
where $C^{\rm tot, s}_{10}$ is defined in Eq.~(\ref{ctot}), and
$C^{\rm tot, d}_{10}$ is given by
\beq 
C^{\rm tot, d}_{10} = C_{10} +
\frac{V_{t^{'}d}V_{t^{'}b}^{*}}{V_{td}V_{tb}^{*}} C_{10}^{t'}\,.
\eeq 
In order to include ${\cal B}(B_q \to \mu^+ \,\mu^-)$ $(q=s,d)$ in the
fit, we define
\beq
B_{\rm lepq} = \frac{16 \pi^3 {\cal{B}}( B_q \to \mu^+ \,\mu^-)}{G^2_F \alpha^2 M_{B_q} m_\mu^2 f_{bq}^2 \tau_{B_q}
 \sqrt{1 - 4 (m_\mu^2/M_{B_q}^2)}} \,.
\eeq
Using the inputs given in Tables \ref{tab1} and \ref{tab3}, we obtain
\beq
B_{\rm leps, {\rm exp}} = 0.025 \pm 0.006 ~~,~~~~ \hskip 30pt B_{\rm lepd, {\rm exp}} = 0.0048\pm 0.0020 ~.
\eeq
The contribution to $\chi^2_{\rm total}$ from ${\cal B}(\bs \to \mu^+
\,\mu^-)$ and ${\cal B}(\bd \to \mu^+ \,\mu^-)$ is then
\beq
\chi^2_{B_q \to \mu^+ \mu^-} = \Big( \frac{B_{\rm leps} - 0.025}{0.006} \Big)^2 + \Big( \frac{B_{\rm lepd} - 0.0048}{0.0020} \Big)^2\;.
\eeq

%%%%%%%%%%%%%%%%%%%%%%%%%%%%%%%%%%%%%%%%%%%%%%%%%%%%%%%%%%%%%%%%%%%%%%%%%%%%%
\subsection{\bf \boldmath Branching ratio of $B\to \tau \,\bar{\nu}$} 
%%%%%%%%%%%%%%%%%%%%%%%%%%%%%%%%%%%%%%%%%%%%%%%%%%%%%%%%%%%%%%%%%%%%%%%%%%

The branching ratio of $B\to \tau \,\bar{\nu}$ is given by
\beq
{\cal{B}}(B\to \tau \,\bar{\nu}) = \frac{G^2_F  M_{B} m_\tau^2}{8\pi} \left(1- \frac{m_\tau^2}{M^2_{B}}\right)^2 f_{bd}^2 |V_{ub}|^2 \tau_{B^{\pm}}.
\eeq
In order to include ${\cal B}(B\to \tau \,\bar{\nu})$ in the fit, we
define
\beq
B_{\rm Btau-nu} = \frac{8\pi {\cal{B}}(B\to \tau \,\bar{\nu})}{G^2_F  M_{B} m_\tau^2 f_{bd}^2 \tau_{B}
 \sqrt{1 - m_\tau^2/M_{B}^2}}\,.
\eeq
Using the inputs given in Tables \ref{tab1} and \ref{tab3}, we obtain
\beq
B_{\rm Btau-nu, {\rm exp}} = (1.779\pm 0.352)\times 10^{-5}.
\eeq
The contribution to $\chi^2_{\rm total}$  from ${\cal B}(B\to \tau \,\bar{\nu})$ is then
\beq
\chi^2_{B \to \tau \,\nu} = \Big( \frac{B_{\rm Btau-nu} - 1.779\times 10^{-5}}{0.352\times 10^{-5}} \Big)^2.
\eeq

%%%%%%%%%%%%%%%%%%%%%%%%%%%%%%%%%%%%%%%%%%%%%%%%%%%%%%%%%%%%%%%%%%
\subsection{\bf \boldmath Like-sign dimuon charge asymmetry $A^b_{SL}$}
%%%%%%%%%%%%%%%%%%%%%%%%%%%%%%%%%%%%%%%%%%%%%%%%%%%%%%%%%%%%%%%%

The (CP-violating) like-sign dimuon charge asymmetry in the $B$ system
is defined as
\bea
 A^b_{SL} \equiv \frac{N_b^{++}  - N_b^{--}}{N_b^{++}  + N_b^{--}} ~,
\eea 
where $N_b^{\pm\pm}$ is the number of events of $b {\bar b} \to
\mu^{\pm} \mu^{\pm} X$. It can be written as
\beq
A^b_{SL} = c^d_{SL} A^d_{SL} + c^s_{SL} A^s_{SL} ~,
\eeq
where $A^q_{SL} = {\rm Im}\Big(\Gamma^{(q)}_{12}/M_{12}^{(q)}\Big)$
$(q=s,d)$, with $c^d_{SL}=0.594 \pm 0.022$ and $c^s_{SL}=0.406 \pm
0.022$. The theoretical expression for $A^q_{SL}$ in the presence of
NP is given in Ref.~\cite{Botella:2014qya}.

$A^b_{sl}$ has been measured by the D\O\ Collaboration. The measured
value is $(-4.96 \pm 1.53 \pm 0.72)\times 10^{-3}$
\cite{Abazov:2013uma}. This deviates by 2.7$\sigma$ from the SM
prediction of $A^b_{SL}$ is $(-2.44 \pm 0.42) \times 10^{-4}$.

The quantities $a$, $b$ and $c$ appear in the theoretical expressions
for $A^q_{SL}$ \cite{Botella:2014qya}. In computing the contribution
to $ \chi^2$ from $A^b_{SL}$, one must include the errors in these
quantities, as well as those in $c^d_{SL}$ and $c^s_{SL}$. To do so,
we consider all of these as parameters and add a contribution to
$\chi^2_{\rm total}$. To be precise,
\beq
\chi^2_{A^b_{SL}} = \Big( \frac{A^b_{SL} - (-4.96\times 10^{-3})}{1.69\times 10^{-3}} \Big)^2 + \chi^2_{c},
\eeq
where
\bea
\chi^2_{c} &=&  \Big( \frac{c^d_{SL} - 0.594}{0.022} \Big)^2 +\Big( \frac{c^s_{SL} - 0.406}{0.022} \Big)^2 \nn\\
&& +~\Big( \frac{a - 10.5}{1.8} \Big)^2 + \Big( \frac{b - 0.2}{0.1} \Big)^2 
+ \Big( \frac{c - (-53.3)}{12} \Big)^2 ~.
\eea

%%%%%%%%%%%%%%%%%%%%%%%%%%%%%%%%%%%%%%%%%%%%%%%%%%%%%%%%%%%%%%%%%%
\subsection{\bf \boldmath The oblique parameter $S$ and $T$}
%%%%%%%%%%%%%%%%%%%%%%%%%%%%%%%%%%%%%%%%%%%%%%%%%%%%%%%%%%%%%%%%%

The theoretical expressions for the oblique parameters $S$ and $T$ in
the VuQ model are given in Ref.~\cite{AgSaav}. For these
non-decoupling corrections we define
\beq
\chi^2_{\rm Oblique} = \Big( \frac{S-0.0}{0.11} \Big)^2 + \Big( \frac{T-0.02}{0.12} \Big)^2 ~.
\eeq

%%%%%%%%%%%%%%%%%%%%%%%%%%%%%%%%%%%%%%%
\section{\bf Results of the fit}
\label{res-fit}
%%%%%%%%%%%%%%%%%%%%%%%%%%%%%%%%%%%%%%%

We first perform a $ \chi^2$ fit to obtain the Wolfenstein parameters
of the standard CKM matrix.  We then redo the fit, using the
theoretical expressions of the VuQ model for the observables. We
obtain values for the Wolfenstein parameters, as well as for the NP
magnitudes $P$, $Q$ and $r$ and the NP phases $\delta_{t'd}$ and $
\delta_{t's}$.  The results of both fits are presented in Table
\ref{table:parameters}, for $m_{t'}=800$ GeV and $1200$ GeV.

%%%%%%%%%%%%%%%%%%%%%%%%%%%%%%%%%%%%%%%%%%%%%%%%%%%%

\begin{table}
\begin{center}
\begin{tabular}{|c|c|c|c|}
\hline
Parameter & SM& $m_{t'}$= 800 GeV&$m_{t'}$= 1200 GeV\\
\hline
$ \lambda$ & $0.226 \pm 0.001$  & $0.226 \pm 0.001$& $0.226 \pm 0.001$ \\
$A$ & $0.780 \pm 0.015$  & $0.770 \pm 0.019$ &$0.769 \pm 0.019$\\
$C$ & $0.39\pm 0.01$  & $0.44\pm 0.02$ & $0.43\pm 0.02$ \\
$ \delta_{ub}$ & $1.21\pm 0.08$  &  $1.13\pm 0.11$& $1.15\pm 0.09$ \\
\hline
$P$ & --  & $0.40 \pm 0.26$ &$0.30 \pm 0.21$ \\
%\label{qvalue}
$Q$ & -- & $0.04 \pm 0.06$&$0.03 \pm 0.05$ \\
%\label{rvalue}
$r$ &  -- & $0.45 \pm 0.25$ &$0.36\pm 0.22$\\
$ \delta_{t'd}$& --  & $0.55 \pm 0.45$ &$0.76 \pm 0.42$\\ 
$ \delta_{t's}$ & --  &  $0.52 \pm 3.26$ &$0.96 \pm 1.21$ \\
\hline
$ \chi^2/d.o.f.$ & $71.15/60$& $63.35/59$& $63.60/59$ \\
\hline
\end{tabular}
\caption{The results of the fits to the parameters of the CKM matrix
  in the SM and in the VuQ model.}
\label{table:parameters}
\end{center}
\end{table}

\begin{table}
\begin{center}
\begin{tabular}{|c|c|c|c|}
\hline
Quantity & SM  & $m_{t'}$= 800 GeV &$m_{t'}$= 1200 GeV\\
\hline
$|V_{ud}|$ & $0.9745 \pm 0.0002$  & $0.9745 \pm 0.0002$&$0.9745 \pm 0.0002$ \\
$|V_{us}|$ & $0.226 \pm 0.001$  & $0.226 \pm 0.001$ &$0.226 \pm 0.001$\\
$|V_{ub}|$ &  $(3.52\pm 0.13) \times 10^{-3}$  & $(3.92\pm 0.24) \times 10^{-3}$&$(3.85\pm 0.21) \times 10^{-3}$ \\
\hline
$|V_{cd}|$ & $0.226 \pm 0.001$  & $0.226 \pm 0.001$&$0.226 \pm 0.001$ \\
$|V_{cs}|$ &  $0.9745 \pm 0.0002$   &  $0.9745 \pm 0.0002$&$0.9745 \pm 0.0002$ \\
$|V_{cb}|$ &  $0.040 \pm 0.001$   &  $0.039 \pm 0.001$ &$0.039 \pm 0.001$\\
\hline
$|V_{td}|$ & $0.0084 \pm 0.0003$ &  $0.0078 \pm 0.0005$ &$0.0080 \pm 0.0004$\\
$|V_{ts}|$ & $0.039 \pm 0.001$  &  $0.039 \pm 0.001$ & $0.039 \pm 0.001$ \\
$|V_{tb}|$ &1   &  $0.995 \pm 0.006$&$0.997\pm 0.004$ \\
\hline
$|V_{t'd}|$ & --  &  $0.005\pm 0.003$ &$0.003\pm 0.002$\\
$|V_{t's}|$ & --    &  $0.002\pm 0.003$& $0.001\pm 0.002$ \\
$|V_{t'b}|$ & --   &  $0.101\pm 0.056$ & $0.082\pm 0.049$\\
%\label{lastV}
\hline
\end{tabular}
\caption{Magnitudes of the $4\times 3$ CKM matrix elements obtained from the fit.}
\label{table:ckm}
\end{center}
\end{table}

%%%%%%%%%%%%%%%%%%%%%%%%%%%%%%%%%%%%%%%%%%%%%%%%%%%%%%%%%%%%%%%%%%%%%%

\begin{table}
\begin{center}
\begin{tabular}{|c|c|c|c|}
\hline
    
Quantity  &SM&  $m_{t'}$= 800 ${\rm GeV}$ &$m_{t'}$= 1200 ${\rm GeV}$ \\
\hline
$|V_{td}V_{tb}^{*}|$ &$0.0084\pm 0.0003$ & $ 0.0077\pm 0.0006$& $ 0.0079\pm 0.0004$ \\
\hline
$|V_{ts}V_{tb}^{*}|$ &$0.0391\pm 0.0008$ &$0.0387 \pm 0.0011$ & $0.0386 \pm 0.001$\\
\hline
$|V_{td}V_{ts}^{*}|$& $ (0.33 \pm 0.02)\times 10^{-3}$&$ (0.30 \pm 0.02)\times 10^{-3}$& $ (0.30 \pm 0.02)\times 10^{-3}$ \\
\hline
$|V_{t'd}V_{t'b}^{*}|$&--  & $ (0.47 \pm 0.40)\times 10^{-3}$& $ (0.28 \pm 0.26)\times 10^{-3}$ \\
\hline
$|V_{t's}V_{t'b}^{*}|$& --  &$ (0.19 \pm 0.32)\times 10^{-3}$ & $ (0.12 \pm 0.20)\times 10^{-3}$\\
\hline
$|V_{t'd}V_{t's}^{*}|$& -- &$ (0.09 \pm 0.15)\times 10^{-4}$& $ (0.05 \pm 0.09)\times 10^{-4}$ \\
\hline
\end{tabular}
\caption{In the VuQ model, combinations of CKM matrix elements that
  control mixing and decay in the $B_d$, $B_s$ and $K$ sectors.
\label{pred:lam}}
\end{center}
\end{table}

%%%%%%%%%%%%%%%%%%%%%%%%%%%%%%%%%%%%%%%%%%%%%%%%%%%%%%%%

From Table \ref{table:parameters}, it can be seen that the
three-generation CKM parameters are not much affected by the addition
of a vector isosinglet up-type quark $t'$. The allowed parameter space
for $C$ and $\delta_{ub}$ expands a little as the constraints on
$|V_{ub}|$ coming from the unitarity of the $3 \times 3$ CKM matrix are
relaxed by the addition of the $t'$ quark.  The new real parameters,
$P$, $Q$ and $r$, are consistent with zero. In addition, the vanishing
of $P$ and $Q$ implies vanishing $V_{t'd}$ and $V_{t's}$,
respectively.  In this case, the phases of these two elements have no
significance.

The magnitudes of the elements of the $4 \times 3$ CKM matrix,
obtained using the fit values of Table~\ref{table:parameters}, are
given in Table~\ref{table:ckm}. From this Table, we find that
$|V_{tb}| \geq 0.98$ at 3$\sigma$. Now, the direct measurement of
$|V_{tb}|$, without assuming unitarity, has been performed using the
single-top-quark production cross section. At the TeVatron one finds
$|V_{tb}|=1.03\pm 0.06$ \cite{Abazov:2013qka,cdf-singletop}, while the
LHC finds $|V_{tb}|=1.03\pm 0.05$ \cite{Chatrchyan:2012ep,Aad:2012ux}.
We therefore see that, although the present direct measurement of
$|V_{tb}|$ is consistent with the SM, a sizeable deviation from its SM
value of 1 is not ruled out due to large experimental errors. On the
other hand, we see that the constraints from present flavor-physics
data do not allow such a sizeable deviation.  We also find that the
allowed values of all of the NP elements of the CKM matrix are
consistent with zero. Furthermore, the 3$\sigma$ upper limits on these
are $|V_{t'd}| \leq 0.01$, $|V_{t's}| \leq 0.01$ and $|V_{t'b}| \leq
0.27$, indicating that the mixing of $t'$ quark with the other three
quarks is constrained to be small.

The values of the magnitudes of the CKM factors that control mixing
and decay in the $B_d$, $B_s$ and $K$ sectors are given in Table
\ref{pred:lam}. In the $b \to s$ sector, the NP contribution is
proportional to the CKM factor $V_{t's}V_{t'b}^{*}$. The corresponding
CKM factor in the SM is $V_{ts}V_{tb}^{*}$. The fit indicates that
$|V_{t's}V_{t'b}^{*}| \ll |V_{ts}V_{tb}^{*}|$. Thus, the NP
contribution in the $b \to s$ sector is tightly constrained in the VuQ
model -- large deviations from the SM predictions are not
possible. This can be seen, for example, from the study of the $B \to
K^* \, \mu^+ \,\mu^-$ observable $P'_5$ in the bin 4.3-8.68 $\rm
GeV^2$ (see Table \ref{bkstar}). The disagreement between the
experimental measurement of $P'_5$ in this bin and its SM prediction
is around the 4$\sigma$ level.  In the SM fit, the $\chi^2_{P'_5}$
contribution to the total $\chi^2_{\rm min}$ is 16.73, reflecting the
large discrepancy between measurement and prediction. In the VuQ fit,
we find $\chi^2_{P'_5}=18.18$ for $m_{t'}=800$ GeV
($\chi^2_{P'_5}=17.36$ for $m_{t'}=1200$ GeV), which shows no
improvement over the SM.

The situation is almost the same in the $b \to d$ and $s \to d$
sectors.  It can be seen from Table \ref{pred:lam} that both
$|V_{t'd}V_{t'b}^{*}|/|V_{td}V_{tb}^{*}|$ and
$|V_{t'd}V_{t's}^{*}|/|V_{td}V_{ts}^{*}|$ are of ${\cal O}(10^{-1})$.
Thus the NP contributions in these sectors from the VuQ model are also
expected to be small.

%%%%%%%%%%%%%%%%%%%%%%%%%%%%%%%%%%%%%%%%%%%%%%%%%%%%%%%%%%%%%%%
\section{\bf \boldmath Predictions for other flavor-physics observables.}
\label{pred:flavor}.
%%%%%%%%%%%%%%%%%%%%%%%%%%%%%%%%%%%%%%%%%%%%%%%%%%%%%%%%%%%%%%%

With the constraints found in the previous section for the NP CKM
matrix elements, it is interesting to see whether any large deviations
from the SM are possible in other flavor-physics observables.  In this
section, we provide predictions for some of the observables in the VuQ
model. These are summarized in Table \ref{predictions}.

\begin{table}
\begin{center}
\begin{tabular}{|c|c|c|c|}
\hline
     & \multicolumn{3}{c|}{Predictions} \\
\cline{2-4} Observable   &SM   &  $m_{t'}$= 800 ${\rm GeV}$ &$m_{t'}$= 1200 ${\rm GeV}$ \\
\hline
${\cal B}(K_L \to \pi^0\,\nu\,\bar{\nu})\times 10^{11}$ & $2.48\pm 0.29$ & $3.24\pm 0.74 $& $3.10\pm 0.59 $ \\
\hline
${\cal B}(B \to X_s\,\nu\,\bar{\nu})\times 10^{5}$ & $2.16\pm 0.23$ & $1.94\pm0.44$ & $1.95\pm0.40$\\
\hline
$x_D$ & Unknown & $ \leq 0.08\%$ at 2$\sigma$& $ \leq 0.03\%$ at 2$\sigma$ \\
\hline
${\cal B}(D \to \mu^+ \mu^-)$ & $ \approx 3 \times 10^{-13}$  & $ (4.56 \pm 10.01) \times 10^{-13}$& $ (1.47 \pm 2.98)\times 10^{-13}$ \\
\hline
${\cal B}(t \to u Z)$ & $ \sim 10^{-17}$ &  $ (1.34 \pm 2.19)\times 10^{-7}$ &  $  (0.50 \pm 0.89)\times 10^{-7}$\\
\hline
${\cal B}(t \to c Z)$ & $ \sim 10^{-14}$&  $  (1.03 \pm 2.69)\times 10^{-7}$ &  $  (0.39 \pm 1.01)\times 10^{-7}$\\
\hline
\end{tabular}
\caption{Predictions for observables in the VuQ model.
\label{predictions}}
\end{center}
\end{table}

%%%%%%%%%%%%%%%%%%%%%%%%%%%%%%%%%%%%%%%%%%%%%%%%%%%%%%%%%
\subsection{\bf \boldmath Branching fraction of $K_L \to \pi^0\,\nu\,\bar{\nu}$}
\label{KL-pinunu}
%%%%%%%%%%%%%%%%%%%%%%%%%%%%%%%%%%%%%%%%%%%%%%%%%%%%%%%%%

In the SM, the decay $K_L\to \pi^0\nu \bar{\nu}$ is dominated by the
short-distance loop diagrams with top-quark exchange, while the
contributions due to the $u$ and $c$ quarks may be neglected.  Thus,
the $t'$ quark in the loop may give a significant contribution.  With
the addition of the $t'$, the branching fraction of $K_L\to
\pi^0\nu\bar{\nu}$ can be written as \cite{SAGMN2,Burasetal}
\beq
{\cal B}(K_L\to \pi^0\nu\bar{\nu}) = \kappa_L 
\left({{\rm Im}(V_{td}V^*_{ts}) \over \lambda^5} X(x_t) + 
{{\rm Im}(V_{t'd}V^*_{t's}) \over \lambda^5} X(x_{t'}) \right)^2 \, ,
\label{brkpi0}
\eeq
with 
\beq
\kappa_L = {r_{K_L}\over r_{K^+}}{\tau(K_L)\over\tau(K^+)}\kappa_+ = (2.31 \pm 0.01) \times 10^{-10}\,.
\eeq
The function $X(x)$ ($x \equiv m^2_{t,t'}/M^2_W$), relevant for the
$t$ and $t'$ pieces, is given by
\beq
X(x) = \eta_X X_0(x)\;,
\eeq
where 
\beq
X_0(x) = {x\over 8}\left[- {2 + x \over 1-x} + {3x-6 \over (1-x)^2} \ln{x}\right]\;.
\eeq
Above, $ \eta_X$ is the NLO QCD correction; its value is estimated to
be $0.994$ \cite{Buras:1997fb}. $r_{K+}$ summarizes the
isospin-breaking corrections in relating $K^+\to \pi^+\nu\bar{\nu}$ to
$K^+\to \pi^0 e^+ \nu $, while $r_{K_L}$ summarizes the isospin
breaking corrections in relating $K_L\to \pi^0\nu\bar{\nu}$ to $K^+\to
\pi^0 e^+ \nu$. 

${\cal B}(K_L \to \pi^0 \nu \bar{\nu})$ is a purely CP-violating
quantity, i.e., it vanishes if CP is conserved. Thus, it is sensitive
to non-standard CP-violating phases.  Within the SM, the branching
ratio of $K_L \to \pi^{0} \nu \bar{\nu}$ can be predicted with very
small uncertainties. It is given by {\cite{Buras:2006gb,Brod:2008ss}}
\begin{equation}
{\cal B}(K_L \to \pi^{0} \nu \bar{\nu}) = (2.48 \pm 0.29)\times 10^{-11}.
\end{equation}
The main source of uncertainty in the branching ratio prediction is
the imaginary part of $V_{td}$. Other theoretical uncertainties are
less than 2\%. Experimentally, this decay has yet to be observed. The
present upper bound on its branching ratio is $2.6 \times 10^{-8}$ at
90\% C.L. \cite{Ahn:2009gb}, which is about three orders of magniture
above its SM prediction. Given the constraints on the $4 \times 3$ CKM
matrix, the VuQ calculation predicts ${\cal B}(K_L\to \pi^0 \nu
\bar{\nu})=(3.24\pm 0.74)\times 10^{-11}$ for $m_{t'}=800\, \rm GeV$
($(3.10\pm 0.59)\times 10^{-11}$ for $m_{t'}=1200\, \rm GeV$).  At
2$\sigma$, ${\cal B}(K_L \to \pi^0 \nu \bar{\nu}) \leq 4.72\times
10^{-11}$, indicating that a large enhancement in the branching ratio
is not allowed.

%%%%%%%%%%%%%%%%%%%%%%%%%%%%%%%%%%%%%%%%%%%%%%%%%%%%%%%%%%
\subsection{\bf \boldmath The branching fraction of $B\to X_s \nu \bar{\nu}$}
\label{BR-Xsnunubar}
%%%%%%%%%%%%%%%%%%%%%%%%%%%%%%%%%%%%%%%%%%%%%%%%%%%%%%%%%%

In the SM, the decay $B\to X_s \nu \bar{\nu}$ is dominated by the
$Z^0$ penguin and box diagrams involving top-quark exchange, and is
theoretically clean. Therefore, we expect that any additional
contributions due to a $t'$ in the loop will be easily identifiable.
The branching fraction for $B\to X_s \nu \bar{\nu}$ in the presence of
a $t'$ quark is given by \cite{SAGMN2}
\beq
{\cal B}(B\to X_s \nu \bar{\nu}) = \frac{\alpha^2 \bar\eta {\cal B}(B\to X_c e\bar{\nu})}{2\pi^2 \sin^4\theta_W |V_{cb}|^2 f(\hat{m}_c) \kappa(\hat{m}_c)}
\Big|V^*_{tb}V_{ts}X_0(x_t)\Big|^2 \Big|1 + \frac{V^*_{t'b}V_{t's}}{V^*_{tb}V_{ts}}\frac{X_0(x_{t'})}{X_0(x_t)}\Big|^2\,.
\eeq
The factor $ \bar \eta \approx 0.83$ represents the QCD correction to
the matrix element of the $b\to s\nu{\bar{\nu}}$ transition due to
virtual and bremsstrahlung contributions, $f(\hat{m}_c)$ is the
phase-space factor in ${\cal B}(B\to X_c e\bar{\nu})$, and
$\kappa(\hat{m}_c)$ is the 1-loop QCD correction factor.  The SM
prediction for ${\cal B}(B \to X_s \nu \bar{\nu})$ is $(2.16\pm 0.23)
\times 10^{-5}$, while in the VuQ model this value changes slightly to
$(1.94\pm 0.44)\times 10^{-5}$ for $m_{t'}=800\, \rm GeV$ ($(1.95\pm
0.40)\times 10^{-5}$ for $m_{t'}=1200\, \rm GeV$). Hence a large
enhancement of the branching fraction of $B\to X_s \nu \bar{\nu}$ is
not allowed.

%%%%%%%%%%%%%%%%%%%%%%%%%%%%%%%%%%%%%%%%%%%%%%%%%%%%%%%%
\subsection{\bf \boldmath $D^0$-${\bar D}^0$ mixing}
%%%%%%%%%%%%%%%%%%%%%%%%%%%%%%%%%%%%%%%%%%%%%%%%%%%%%%%%

Within the SM, $D^0$-${\bar D}^0$ mixing occurs at loop level and
involves the lighter quarks $d$, $s$ and $b$. This implies a strong
Glashow-Iliopoulos-Maiani (GIM) cancellation, and hence a small SD
contribution. Furthermore, the $b$-quark contribution is highly
suppressed, $O(\lambda^8)$, so that the mixing is dominated by the
$d$- and $s$-quark contributions.  There are, therefore, large LD
contributions to $D^0$-${\bar D}^0$ mixing, and indeed they dominate
over the SD contributions.  The present measurement of the
$D^0$-${\bar D}^0$ mixing parameter $x_D$ is
\beq 
x_D \equiv \frac{\Delta M_D}{\Gamma_D} = (0.8 \pm 0.1)\%\, .
\eeq 
This is much larger than the short-distance SM prediction. Still, in
order to determine if the SM can explain this value of $x_D$, one must
have an accurate estimate of the LD contribution. Unfortunately, this
is not available at present.

As noted in the introduction, the mixing of the $t'_L$ with $\{u_L,
c_L, t_L\}$ will induce tree-level $Z$-mediated FCNC's among the SM
quarks. Thus, in the VuQ model, $D^0$-${\bar D}^0$ mixing occurs at
tree level. It may therefore provide a much larger contribution than
that of the (short-distance) SM.  Neglecting the SM contributions, in
the VuQ model $D^0$-${\bar D}^0$ mixing is given by
\cite{Golowich:2009ii}
\beq
x_d = \frac{G_F |U_{uc}|^2 f^2_D M_D B_D r(m_c, M_Z)}{3 \sqrt{2}\Gamma_D} ~,
\eeq
where $|U_{uc}|=V_{u4}V^*_{c4}$ is the $Z$-$u$-$c$ flavor-changing
coupling, and $r(m_c, M_Z)=0.778$ is the renormalization-group
factor. Using $f_D= 209.2 \pm 3.3$ MeV \cite{Aoki:2013ldr},
$B_D=1.18\pm 0.07$ \cite{Buras:2010nd} and ${\bar\tau}_D = 0.4101$ ps
\cite{pdg}, we find that, given the constraints on $V_{u4}V^*_{c4}$,
in the VuQ model, $x_D=(0.016 \pm 0.034)\%$ for $m_{t'}=800\, \rm
GeV$ ($(0.005 \pm 0.010)\%$ for $m_{t'}=1200\, \rm GeV$). Thus at
2$\sigma$, $x_D \leq 0.08\%$. We therefore see that the SD
contribution in the VuQ model falls far below the observed value of
$D^0$-${\bar D}^0$ mixing.

%%%%%%%%%%%%%%%%%%%%%%%%%%%%%%%%%%%%%%%%%%%%%%%%%%%%%%%%
\subsection{\bf \boldmath Branching fraction of $D^0 \to \mu^+ \mu^-$}
%%%%%%%%%%%%%%%%%%%%%%%%%%%%%%%%%%%%%%%%%%%%%%%%%%%%%%%%

Unlike $D^0$-${\bar D}^0$ mixing, the SM prediction for the branching
fraction of $D^0 \to \mu^+ \mu^-$ can be estimated fairly accurately,
even after including the LD contribution. The SM prediction for the
$D^0 \to \mu^+ \mu^-$ branching ratio is $ \approx 3 \times 10^{-13}$,
hence highly suppressed. Thus, $D^0 \to \mu^+ \mu^-$ has the potential
for large NP contributions. At present, we only have an experimental
upper bound on the branching ratio: ${\cal B}(D^0 \to \mu^+ \mu^-)\leq
7.6 \times 10^{-9}$ at 95\% C.L. \cite{Aaij:2013cza}, which is several
orders of magnitude larger than the SM prediction.

Within the VuQ model, $D^0 \to \mu^+ \mu^-$ occurs at tree level due
to $Z$-mediated FCNC's. Neglecting the SM contribution, the branching
ratio in the VuQ model is given by \cite{Golowich:2009ii}
\beq
{\cal B}(D^0 \to \mu^+ \mu^-) = \frac{G_F m^2_{\mu} f^2_D M_D}{32 \pi \Gamma_D} \sqrt{1-\frac{4 m^2_{\mu}}{m^2_D}} |U_{uc}|^2 ~.
\eeq
For $m_{t'}=800\, \rm GeV$, ${\cal B}(D^0 \to \mu^+ \mu^-) = (4.56\pm
10.01)\times 10^{-13}$ ($(1.47 \pm 2.98)\times 10^{-13}$ for
$m_{t'}=1200\, \rm GeV$).  Thus, at 2$ \sigma$, ${\cal B}(D^0 \to
\mu^+ \mu^-) \leq 2.46 \times 10^{-12}$. We therefore observe that the
branching ratio of $D^0 \to \mu^+ \mu^-$ can be enhanced by an order
of magnitude above its SM value, but this is still far below the
present detection level.

%%%%%%%%%%%%%%%%%%%%%%%%%%%%%%%%%%%%%%%%%%%%%%%%%%%%%%%%
\subsection{\bf \boldmath Branching fraction of $t\to q Z$ ($q=c,u$)}
%%%%%%%%%%%%%%%%%%%%%%%%%%%%%%%%%%%%%%%%%%%%%%%%%%%%%%%%

Within the SM, the branching ratios of the FCNC top decays $t \to u Z$
and $t \to c Z$ are $ \sim 10^{-17}$ and $ \sim 10^{-14}$,
respectively \cite{Eilam:1990zc,AguilarSaavedra:2004wm}.  The present
upper bound on ${\cal B} (t \to q Z)$ is 0.21\% at 95\%
C.L. \cite{Chatrchyan:2012hqa}.  The discovery potential of ${\cal
  B}(t\to q Z)$ is $ \sim 10^{-4}$-$10^{-5}$ at ATLAS and CMS. The SM
value of ${\cal B} (t \to q Z)$ is thus far below the detection level
for these decays. This implies that these decays can only be observed
if NP enhances their branching ratios by many orders of magnitude
above their SM values.

This may be possible within the VuQ model, as here, due to
$Z$-mediated FCNC's, these decays occur at tree level. Neglecting the
SM contribution, the decay rate for $t\to q Z$ is given by
\cite{AguilarSaavedra:2004wm}
\beq
\Gamma(t \to q Z) = \frac{\alpha}{32 \sin^2\theta_W \cos^2\theta_W} |U_{qt}|^2 \frac{m^3_t}{M^2_Z}\Big[1-\frac{M^2_Z}{m^2_t}\Big]^2 \Big[1+2\frac{M^2_Z}{m^2_t}\Big]\,,
\eeq
where $m_t=173.2\pm0.9$ GeV \cite{Aaltonen:2012ra} and
$|U_{qt}|=V_{q4} V^*_{t4}$.  As $V_{tb}$ in this model is close to
unity, we can approximate the top width by $ \Gamma(t \to b W^+)$,
which at leading order is given by
\beq
\Gamma(t \to b W^+)=\frac{\alpha}{16 \sin^2\theta_W} |V_{tb}|^2 \frac{m^3_t}{M^2_W}\Big[1-3\frac{M^4_W}{m^4_t}+2\frac{M^6_W}{m^6_t}\Big]\,.
\eeq
The branching ratio of $t\to q Z$ is therefore given by 
\beq 
{\cal B}(t \to q Z) = (0.463 \pm 0.001) \frac{|U_{qt}|^2}{|V_{tb}|^2}\,.  
\eeq 
Using the values of parameters given in Table~\ref{table:parameters},
we obtain $|U_{ut}|=(0.53 \pm 0.43)\times 10^{-3}$ ($(0.33 \pm 0.29)\times
10^{-3}$) and
$|U_{ct}|=(0.47 \pm 0.61)\times 10^{-3}$ ($(0.29 \pm 0.37)\times
10^{-3}$) for $m_{t'}=$ 800 GeV (1200 GeV). This leads to ${\cal B}(t
\to u Z) = (1.34 \pm 2.19) \times 10^{-7}$ ($(0.50\pm 0.89) \times
10^{-7}$) and ${\cal B}(t \to c Z) =(1.03 \pm 2.69) \times 10^{-7}$
($(0.39 \pm 1.01) \times 10^{-7}$) for $m_{t'}=$ 800 GeV (1200 GeV).
Therefore, the FCNC branching ratios can indeed be enhanced by many
orders of magnitude above their SM values. However, they are still two
orders of magnitude below the present detection level for these
decays.

%%%%%%%%%%%%%%%%%%%%%%%%%%%%%%%%%%%%%%%%%%%%%%%%%%%%
\section{\bf Conclusions}
\label{concl}
%%%%%%%%%%%%%%%%%%%%%%%%%%%%%%%%%%%%%%%%%%%%%%%%%%%%

In this paper we consider the VuQ model, in which a vector isosinglet
up-type quark $t'$ is added to the standard model (SM). In the VuQ
model, the full CKM quark mixing matrix is $4 \times 3$, and is
parametrized by four SM and five new-physics (NP) parameters. The NP
parameters include three magnitudes and two (CP-violating) phases. We
perform a fit using flavor-physics data to constrain all CKM
parameters. The purpose is to determine whether there are any
indications of NP, such as the non-unitarity of the $3 \times 3$ SM
CKM matrix, or, equivalently, nonzero values for some of the NP
parameters. And even if there is no evidence of NP, we would like to
ascertain whether sizeable NP effects are still possible in other
flavor-physics observables, while being consistent with the
constraints found in the fit.

The fit involves 68 flavor-physics observables. 
No evidence for NP is found: the values of the three NP
magnitudes are consistent with zero, in which case the two NP phases
have no significance. Specific results include the following:
\begin{itemize}

\item The deviations of the CKM matrix elements $V_{ts}$ and $V_{td}$
  from their SM prediction are small.

\item At 3$\sigma$, $|V_{tb}| \geq 0.98$. Any large deviation of
  $|V_{tb}|$ from unity is therefore not possible in the VuQ model.

\item The 3$\sigma$ upper limits on the new elements of the VuQ CKM
  matrix are: $|V_{t'd}| \leq 0.01$, $|V_{t's}| \leq 0.01$ and
  $|V_{t'b}| \leq 0.27$, indicating that the mixing of $t'$ quark with
  the other three quarks is constrained to be small.

\end{itemize}

Turning to possible NP effects in the VuQ model, we find that any NP
contributions to $b \to s$, $b \to d$ and $s \to d$ transitions are
tightly constrained. We also find,
\begin{itemize}

\item A large enhancement of SD contribution to $x_d$ (i.e.,
  $D^0$-${\bar D}^0$ mixing) is not allowed.

\item The branching ratio of $D^0 \to \mu^+ \mu^-$ can be enhanced by
  an order of magnitude above its SM value, but this is still  far below
  the present detection level.

\item The branching ratios of the flavor-changing decays $t \to q Z$
  ($q=c,u$) can be enhanced by many orders of magnitude.  However,
  they are still  two orders of magnitude below the present
  detection level.

\end{itemize}

In summary, current flavor data puts extremely stringent constraints
on the VuQ model. There are no hints of NP in the CKM matrix.
Furthermore, the fit to the data indicates that any VuQ contributions
to loop-level flavor-changing $b \to s$, $b \to d$ and $s \to d$
transitions are very small. There can be significant enhancements of
the branching ratios of $t \to u Z$ and $t \to c Z$ decays, but these
are still below detection levels.

\bigskip
\noindent
{\bf Acknowledgments}: DK would like to thank Farvah Mahmoudi,
J. Virto and Sanjeev Kumar for help in the numerical analysis related
to $B \to K^* \mu^+ \mu^-$.  The work of AKA and SB is supported by
CSIR, Government of India, grant no: 03(1255)/12/EMR-II. The work of
DL was financially supported by NSERC of Canada.

\end{document}